# Automatic Census of Mussel Platforms Using Sentinel 2 Images

Fernando Martín-Rodríguez, Fernando Isasi-de-Vicente, Mónica Fernández-Barciela.
fmartin@tsc.uvigo.es, fisasi@tsc.uvigo.es, monica@tsc.uvigo.es.
atlanTTic research center for Telecommunication Technologies, University of Vigo,
Campus Lagoas Marcosende S/N, 36310 Vigo, Spain.

***Abstract*- Mussel platforms are big floating structures made of wood (size is normally about 20x20 meters or even a bit larger) that are used for aquaculture, id EST: growing mussels in appropriate marine waters. These structures are very typical in Galician estuaries. Being interesting to produce a periodic census of these structures that would allow knowing their number and positions, as well as changes on those parameters; Satellites that obtain periodic images for earth observation are a natural election for this issue. This paper describes a preliminary application able to construct automatically such a census using Sentinel 2 images (Copernicus Project). Copernicus satellites are run by European Space Agency (ESA) and the produced images are freely distributed on the internet. Sentinel 2 images have thirteen frequency bands and are updated each five days. In our application, we use remote sensing normalized (differential) indexes and artificial Neural Networks applied to multiband data. Different methods are described and tested. Finally, results are presented.**

## I. Introduction

Mussel platforms (or rafts) are large floating structures made of wood. Their size is usually about 20x20 meters or even a bit more. They that are used for aquaculture (growing mussels). These structures are typical of the Galician estuaries, being interesting to elaborate a periodic census of these structures that would allow to know their number and positions, as well as to detect changes, new and decommissioned ones... Satellites that obtain periodic images for Earth observation are a natural choice for this topic. In [1] they work with the same purpose, using SAR (Synthetic Aperture Radar) data. Our article describes an application that, conversely, uses the optical information of Sentinel 2. Despite of being a preliminary version, we can automatically construct the desired platform census. Sentinel 2 is a satellite of the Copernicus project [2], operated by the European Space Agency (ESA). The produced images are freely obtained on the Internet [3], these are multispectral images of thirteen bands [4] (figure 1) that are updated every five days. In our application, we use normalized differential indices (very typical in remote sensing) and also artificial Neural Networks applied to multiband data. Different methods are described and tested and the results are presented.

The images of Sentinel 2 have a spatial resolution of 10 meters per pixel (really, only some bands have that pixel size, there exist bands of 20 m and 60 m) which would result in a size of 2x2 or 3x3 for the rafts. In addition, as the structure of a platform is not a continuous wooden platform but, rather, a lattice of planks (figure 2); the rafts appear in the visible bands only as small squares within the water with a color a little less saturated than their surroundings (figure 3). Therefore, we will need to use the non-visible bands of the image to be able to make reliable detections.

Sentinel 2 only has coverage in near-shore waters and inland seas. In our case, this is more than enough.

| Sentinel-2 Bands | Central Wavelength (µm) | Resolution (m) |
|---|---|---|
| Band 1 - Coastal aerosol | 0.443 | 60 |
| Band 2 - Blue | 0.490 | 10 |
| Band 3 - Green | 0.560 | 10 |
| Band 4 - Red | 0.665 | 10 |
| Band 5 - Vegetation Red Edge | 0.705 | 20 |
| Band 6 - Vegetation Red Edge | 0.740 | 20 |
| Band 7 - Vegetation Red Edge | 0.783 | 20 |
| Band 8 - NIR | 0.842 | 10 |
| Band 8A - Vegetation Red Edge | 0.865 | 20 |
| Band 9 - Water vapour | 0.945 | 60 |
| Band 10 - SWIR - Cirrus | 1.375 | 60 |
| Band 11 - SWIR | 1.610 | 20 |
| Band 12 - SWIR | 2.190 | 20 |

Fig. 1. Sentinel Bands 2.

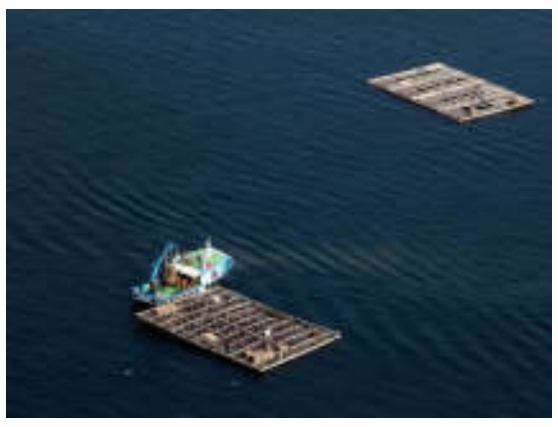

Fig. 2. Bateas from the air.

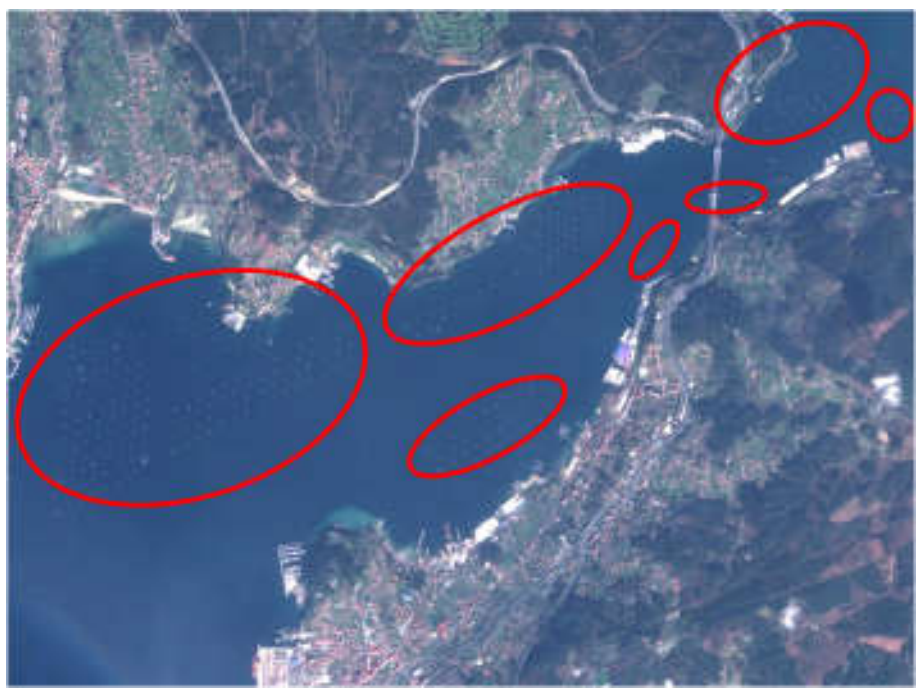

Fig. 3. Polygons of rafts in image Sentinel 2 (Vigo estuary).

Sentinel's public repository contains images of 100x100 Km (100 Mpx with 10m pixels) that comprise all bands and are updated every five days. For each image, we have two versions: the TOA correction that contains the thirteen bands and the BOA correction that only contains twelve since the band ten is used within the correction process to estimate the atmospheric state [5].

We have tested our system with both types of images. In both cases, we have discarded the 60 m bands because of the excessive scaling that we would need to do to combine them with the others and because they provide information very dependent on the atmosphere.

## II. Phase One: Water Detection

Our first objective is to detect an area of interest where to apply a detector that can distinguish the points belonging to platforms. We could use a map instead to work always over sea points, but we do not have such kind of maps. In addition, a water detection method will eliminate cloud areas and will also consider the effect of tides.

### A. *Detection by Normalized Indexes*

In remote sensing, the so-called normalized indexes are used very often. Index are calculated from pairs of components [6,7]. In particular, the NDWI (Normalized Differential Water Index), is defined as:

$$NDWI = \frac{GREEN-NIR}{GREEN+NIR} \quad (1)$$

This value is calculated from bands 3 (GREEN) and 8 (NIR). NDWI will always be in the range [-1, +1], the positive values will tend to correspond to bodies of water, while the negative ones will be dry areas.

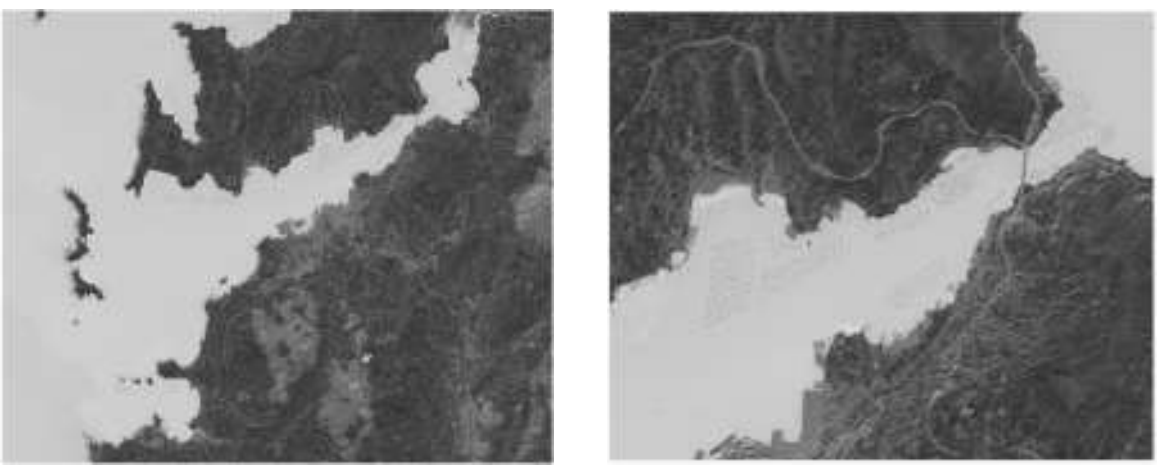

Fig. 4. NDWI represented in grayscale.

As we can see in Figure 4, the brighter (numerically larger) values correspond to water. However, the value obtained for water is different in images of different days. By making all negative pixels equal to zero, a bimodal histogram is achieved with a strong peak at zero and another one corresponding to the water regions.

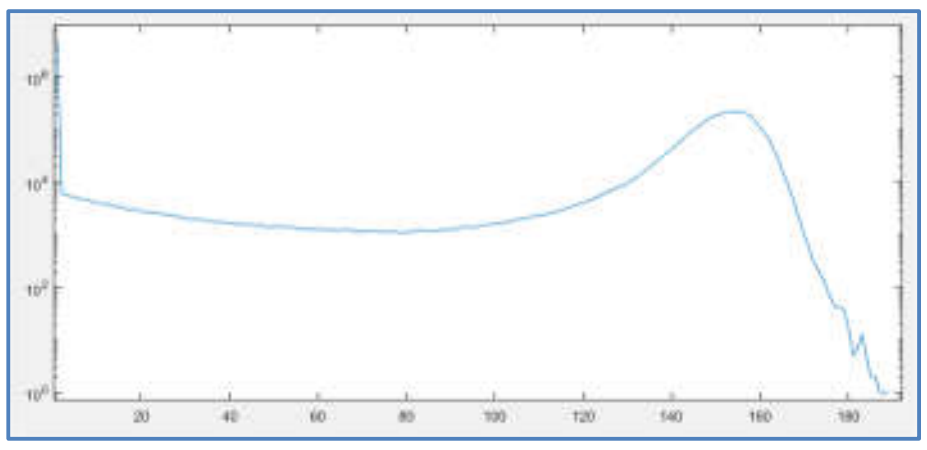

Fig. 5. Histogram of the NDWI image (logarithmic scale).

At this time, the well-known Otsu method [8] will allow us to calculate an adequate threshold to distinguish water.

### B. *Detection using neural networks*

Using the same methodology as in [9], if we define a vector of characteristics for each pixel consisting of the values of each band at that point, we will have a numerical vector of size 10 (we have eliminated the lower resolution bands: 1, 9 and 10). Note that for the bands of resolution equal to 20 m we will have to perform an interpolation, for which we choose the filter "Lanczos3" [10]. To classify these vectors we train a simple neural network of the MLP (Multi-Layer-Perceptron) type [11].

In this case we have trained the network to distinguish 5 types of surfaces: 1-> empty (part of image without information), 2-> solid ground, 3-> water, 4-> cloud and 5-> foam (foam points on the coastline, very typical in the Atlantic).

The structure of the network is typical of MLPs: 10 inputs (size of the characteristic vector), 5 outputs (one for each class) and 10 hidden neurons (a number adjusted empirically).

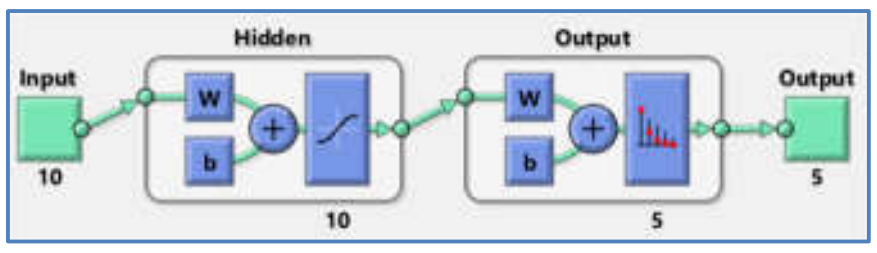

Fig. 6. Network structure.

The training has been carried out with labeled points obtained from real images. The number of samples per class has been balanced by preserving the water samples (class of interest) and randomly eliminating samples from the majority classes. The training method has been "Backpropagation with conjugate gradient" [12] and the computer tool MATLAB [13].

The results have been good for all classes except foam. We can see them in Figure 7 (confusion matrices).. 70% of the samples were used for training, 15% to validate and finish the training and the remaining 15% for the final test (test). The total number of samples is greater than 19 million.

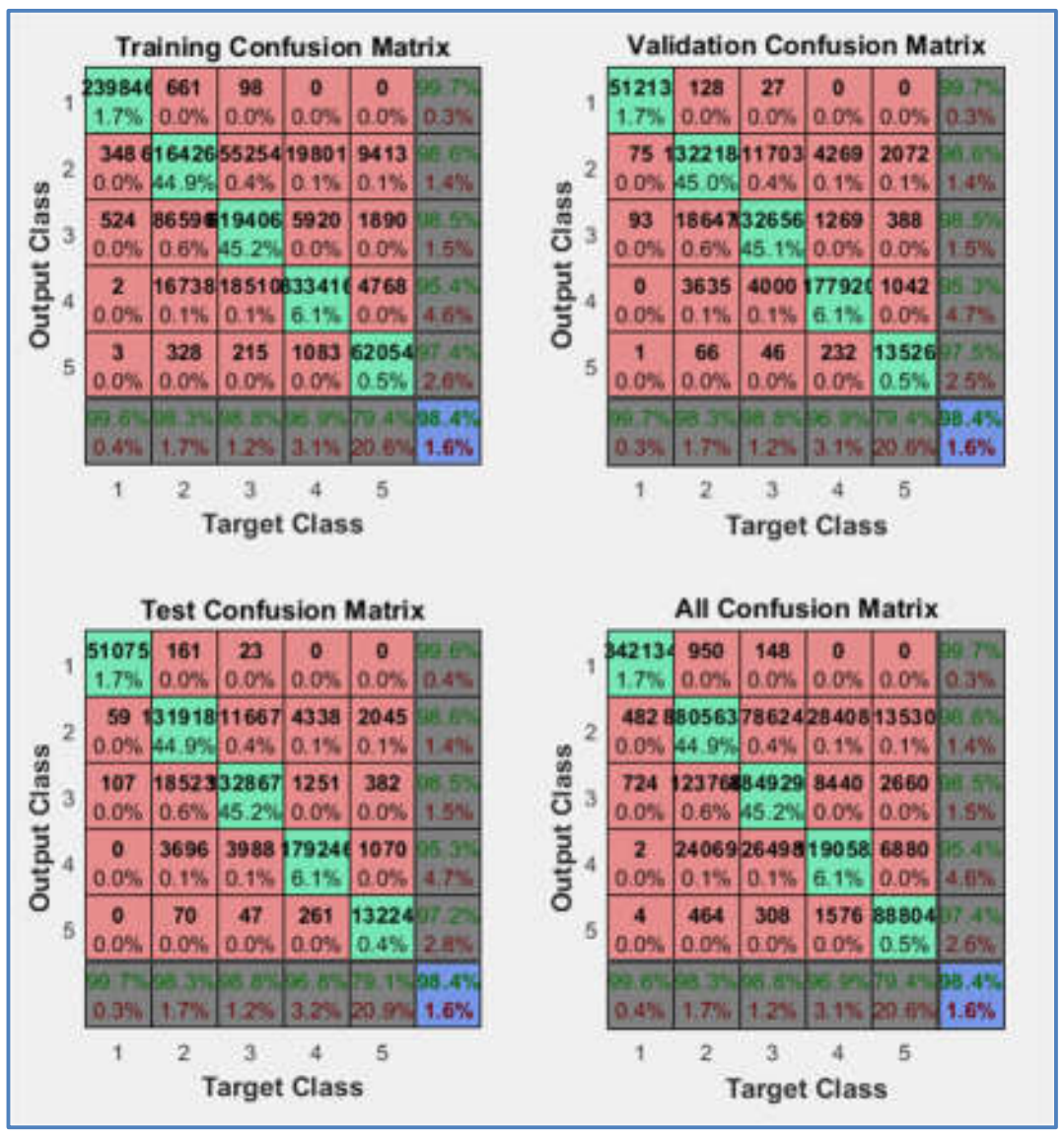

Fig. 7. Confusion Matrices.

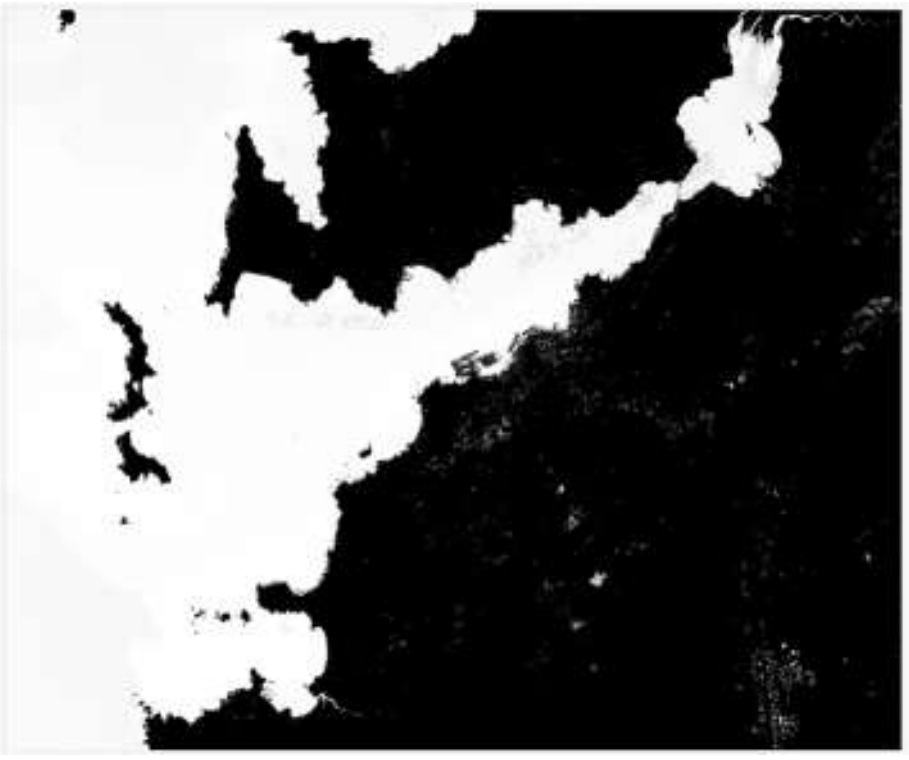

Fig. 8. Detection with neural networks.

In figure 8 we see the result obtained for a sub-image containing the Vigo estuary where the output 3 (water) of the neural network has been represented as an image. Values close to 1.0 mean positive water detection. The obtained mask is processed using mathematical morphology [14] to be cleaned and compacted. Process is (expressed in mathematical morphology terms): 1 -> closing, 2-> opening and 3-> erosion (used to eliminate points very close to the coastline). These same operations are also performed with the mask obtained by the alternate method (NDWI). The threshold for binarizing the output of the neural network (a number between 0.0 and 1.0) is 0.90.

## III. Detection of Platforms

Now it is the question of classifying all the pixels previously detected as water (those that have a positive value in the masks obtained in the previous section). The result of this classifier will be binary: "platform" or "not platform". This classifier is based on a second neural network. The results obtained are treated as an image of connected components (blobs) that are possible platforms. This image is processed by mathematical morphology in order to eliminate false positives that would reduce the final success rate.

### *A. Neural Network*

In this case, we use an MLP again. Now we have ten inputs again (ten bands of sufficient resolution) and a single output neuron (the value obtained will be close to 1.0 when we are detecting a platform). For this second case, we can use fewer neurons at the intermediate level: in particular, we have achieved training convergence with only two hidden neurons (figure 9).

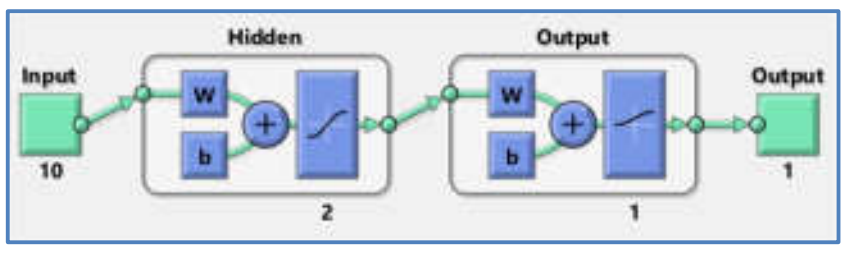

Fig. 9. Network structure.

As we can see in figures 4 and 8, water masks usually present dark holes in the platform points. Obviously, this is a negative detection, that is: "it happens because those points are NOT water". When processing the mask, the closing operation makes those and other holes (holes due to boats or other floating objects) disappear. A morphological operation known as "Bottom Hat" (or negative "Top Hat") would allow us to obtain those points as active points (white) on a black background: `BottomHat(Im) = Close(Im)-Im`. That wouldn't be a detection of enough reliability. Nevertheless, we use this method (manually corrected) to find training samples.

The training has been carried out with the same method that we explained in the previous section. The total number of samples is 12976. It has been based on 6488 samples (pixels) of platforms in sub-images of the estuaries of Pontevedra and Vigo. Afterward, the same number of water samples have been obtained, randomly extracted from the same images.

In Figure 10 we present the confusion matrices for this new network where it is shown that the error rate is below 2% in all cases.

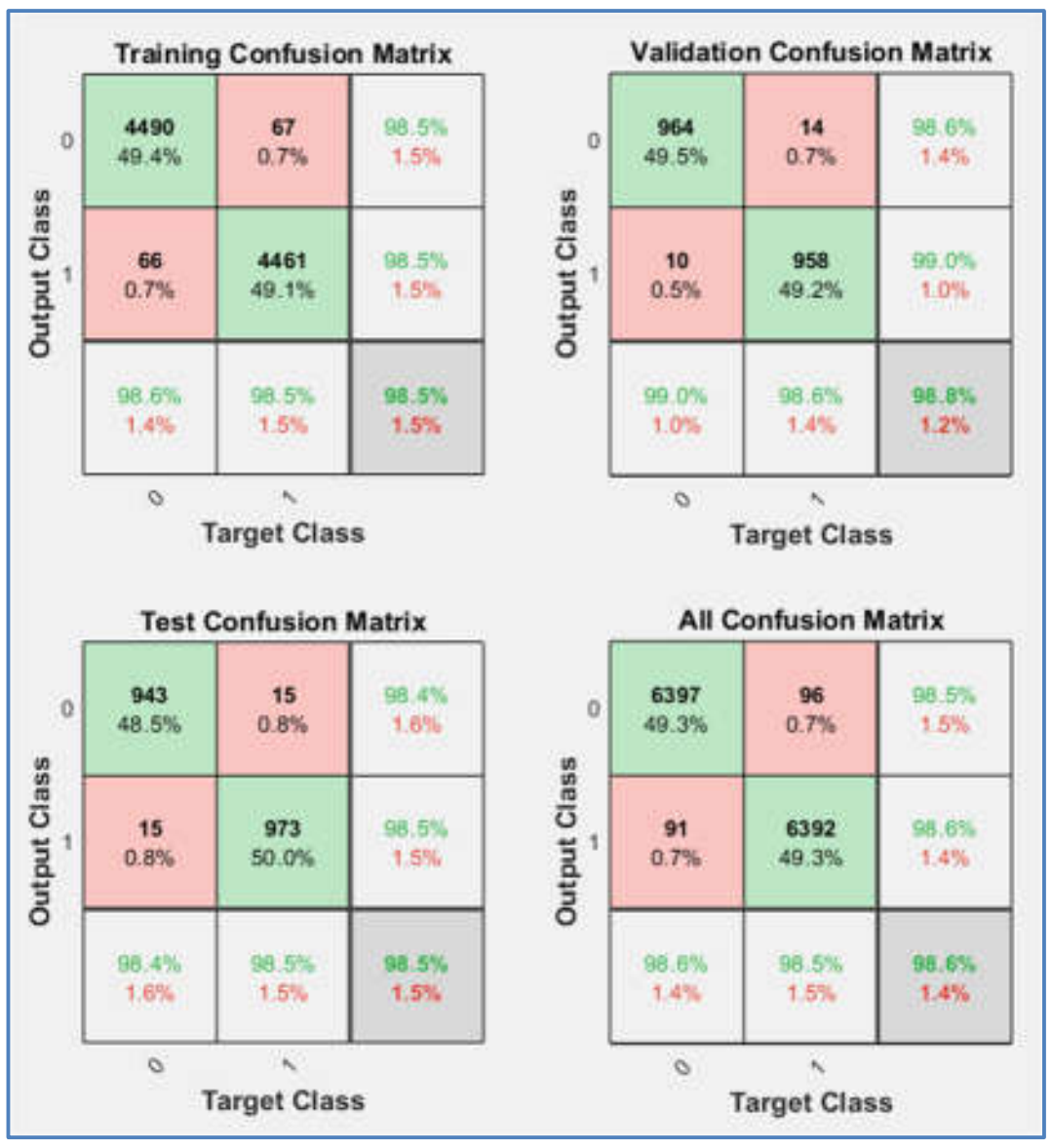

Fig. 10. Confusion Matrices.

### *B. Post-processing of the results*

The results on other images of the same estuaries and, also on other estuaries, were good; but some false positives were detected on other man-made structures. As an example (figure 11), we see a false positive on a bridge in the Estuary of Noia (besides the bridge, two ancient stone structures result in other, line shaped, false positive blob).

These types of errors can be easily eliminated according to its irregular shape and its size much larger than a platform.

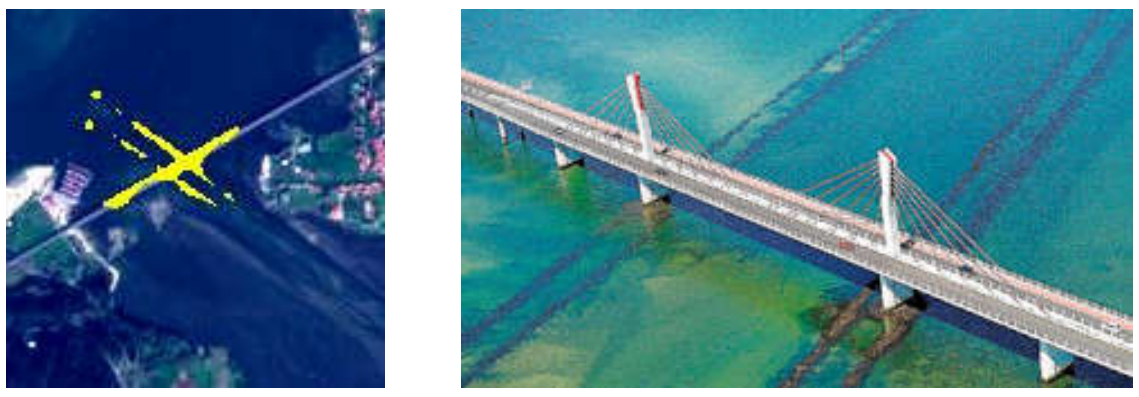

Fig. 11. Example of false positive (bridge of the Ría de Noia).

Therefore, the output of the neural network (only active) on the water mask) is post-processed. For each connected object (blob), conditions are imposed on its geometry: "area less than a maximum", "equivalent diameter less than a maximum", "Euler number equal to 1" (number of components minus number of gaps) and "solidity greater than a minimum" (percentage of blob points versus the area of the "ConvexHull"). With this filtering, highly satisfactory results are obtained.

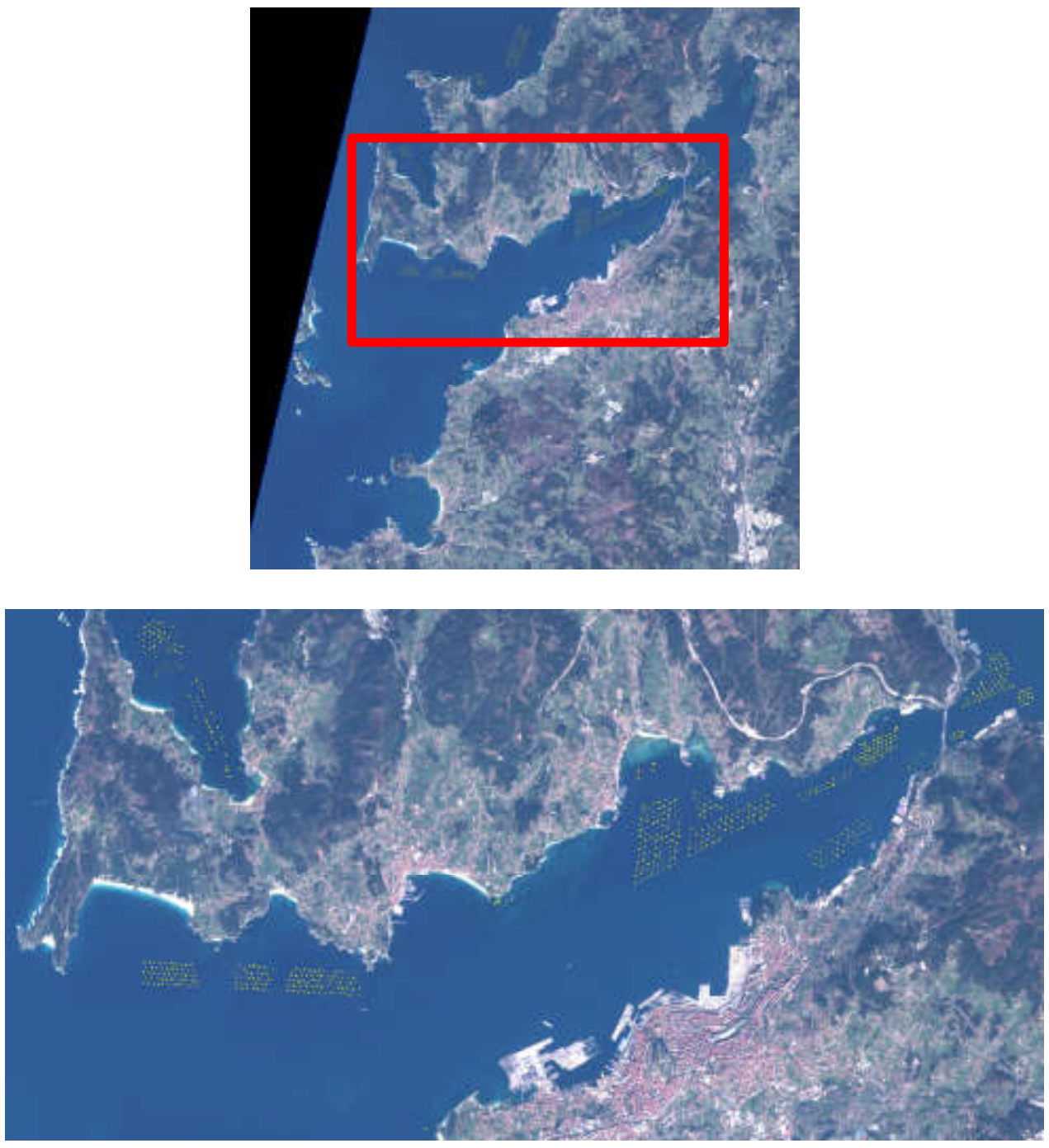

Fig. 12. Example of result (Vigo estuary).

## IV. RESULTS

We have applied the study to Sentinel image clippings corresponding to the estuaries of Vigo (635 platforms), Pontevedra (321 platforms), Arousa (2307 platforms), Noia (126 platforms) and Corcubión (it is not a suitable area and there are no platforms, it has been added as a control image). Images from different dates have been used. As it is due in supervised learning, the images used for training have NOT been included in tests. The results obtained are summarized in the following tables:

TABLA I
TOA correction (without 60M bands), NDWI for water detection.

| **TFA (False Acceptance Rate)** | **TFR (False Rejection Rate)** |
|---|---|
| 8.54% | 0.82% |

TABLA II
TOA correction (without 60M bands), MLP for water detection.

| **TFA (False Acceptance Rate)** | **TFR (False Rejection Rate)** |
|---|---|
| 8.97% | 0.88% |

Where TFA and TFR are defined as:

$$TFA = \frac{FalseDetections}{TotalDetections} 100$$

$$TFR = \frac{MissedDetections}{TotalPlatforms} 100 \quad (2)$$

For images with BOA correction, many false positives have been observed, very difficult to eliminate, which, at least for the moment, make this option a bad choice (figure 13).

## V. CONCLUSIONS AND FUTURE LINES

We have developed a method capable of locating the mussel platforms of the Galician estuaries (that can be used anywhere else), using Sentinel 2 images and MATLAB processing (which, of course, can be implemented over other platforms).

For this particular problem, it seems better to use images with TOA correction (L1C) than those with BOA correction (L2A).

Between the two methods used to detect water bodies (NDWI and MLP), the results of Tables I and II recommend the NDWI-based method.

As future lines we would highlight:

- Process automation, implementing it in an environment more suitable for a end-user application (C ++ or pyhton), performing the automatic download and cropping of the images.
- Study of the reasons that make worse the results with BOA correction.

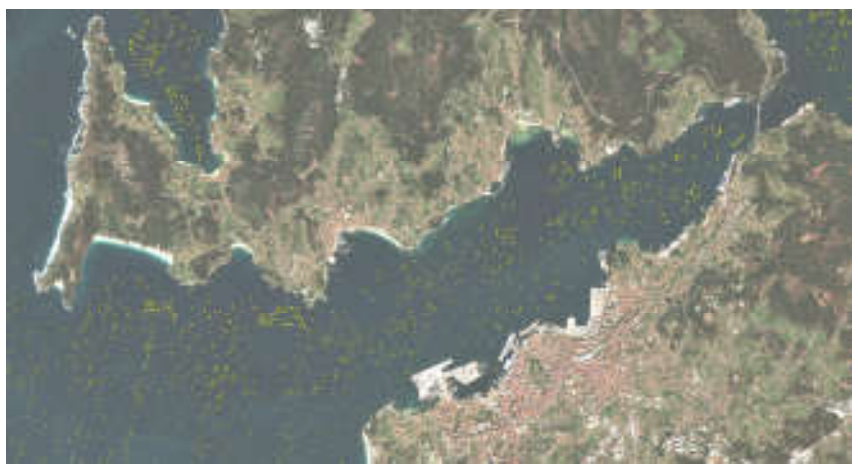

Fig. 13. Use of BOA correction (Pontevedra estuary).

## REFERENCES

[1] A. Marino et al, "Detecting aquaculture platforms using COSMO SkyMed", EUSAR 2021 (13th European Conference on Synthetic Aperture Radar, IEEE), 2021.

[2] https://www.copernicus.eu/en/about-copernicus/infrastructure/discover-our-satellites, access: 27/03/2022.

[2] https://scihub.copernicus.eu/, accessed: 27/03/2022.

[3] https://sentinel.esa.int/web/sentinel/user-guides/sentinel-2-msi/resolutions/spatial, access 27/03/2022.

[4] https://inta.es/INTA/es/blogs/copernicus/BlogEntry_1509095468013#, accessed: 27/03/2022.

[5] B. Gao, "NDWI A Normalized Difference Water Index for Remote Sensing of Vegetation Liquid Water From Space", *Remote Sensing of the Environment* (Elsevier), vol 58, 257-266, 1996.

[6] S.K. McFeeters, "The use of the Normalized Difference Water Index (NDWI) in the delineation of open water features", *International Journal of Remote Sensing*, Volume 17, 1996 - Issue 7.

[7] N. Otsu, "A threshold selection method from gray level histograms", IEEE Transactions on Systems, Man, and Cybernetics, No. 9(1), pp62-66, 1979.

[8] F. Martín-Rodríguez, O. Mojón-Ojea, "Big plastic masses detection using satellite images & machine learning", Instrumentation Viewpoint, Nº 21, pp30-31, 2021.

[9] P. Getreuer, "Linear Methods for Image Interpolation", *Image Processing On Line*, vol 1, pp. 238–259 (2011).

[10] S. Haykin, *Neural Networks: A Comprehensive Foundation (2 Edition)*, Upper Saddle River, NJ (U.S.): Prentice Hall, 1998.

[11] E.M. Johansson et al, "Backpropagation Learning for Multilayer Feed-Forward Learning Neural Networks Using the Conjugate Gradient Method", International Journal of Neural Systems, Vol. 02, No. 04, pp. 291-301 (1991).

[12] MATLAB, https://es.mathworks.com/products/matlab.html. Last accessed 27/03/2022.

[13] R.C. González et al., "Digital image processing using MATLAB", Upper Saddle River : Gatesmark Publishing, cop. 2009.